\documentclass[conference]{IEEEtran}
\ifCLASSINFOpdf
   \usepackage[pdftex]{graphicx}
\else
   \usepackage[dvips]{graphicx}
\fi
\usepackage{amssymb}
\usepackage{amsmath}
\usepackage{booktabs}
\usepackage{mdwlist}

\begin{document}
%
\title{Efficient and Spontaneous Privacy-Preserving Protocol for Secure Vehicular Communications}

\author{Hu Xiong$^{\dag}$$^{\ddag}$, Matei Ripeanu$^{\ddag}$, Zhiguang Qin$^{\dag}$
\\$^{\dag}$School of Computer Science and Engineering,
\\University of Electronic Science and Technology of China, Chengdu, P.R. China
\\$^{\ddag}$Department of Electrical and Computer Engineering, \\The University of British Columbia, Vancouver, BC, Canada
\\Email: xionghu.uestc@gmail.com, matei@ece.ubc.ca,
qinzg@uestc.edu.cn }


%


\maketitle

\begin{abstract}

This paper introduces an efficient and spontaneous
privacy-preserving protocol for vehicular ad-hoc networks based on
revocable ring signature. The proposed protocol has three appealing
characteristics: First, it offers \textit{conditional
privacy-preservation}: while a receiver can verify that a message
issuer is an authorized participant in the system only a trusted
authority can reveal the true identity of a message sender. Second,
it is \textit{spontaneous}: safety messages can be authenticated
locally, without support from the roadside units or contacting other
vehicles. Third, it is \textit{efficient} by offering fast message
authentication and verification, cost-effective identity tracking in
case of a dispute, and low storage requirements. We use extensive
analysis to demonstrate the merits of the proposed protocol and to
contrast it with previously proposed solutions.

\end{abstract}

%
\IEEEpeerreviewmaketitle
\section{Introduction}

Each year, over six million crashes occur on U.S. highways. These
accidents kill more than 42,000 people, injure three million others,
and cost more than \$230 billion per year\cite{IVI}. To reduce the
number and the severity of these crashes and to improve driving
experience, car manufactures and the telecommunication industry
recently have geared up to equip each vehicle with wireless devices
that allow vehicles to communicate with each other as well as with
the roadside infrastructure\cite{VII,Misener2005}. These wireless
communication devices installed on vehicles, also known as onboard
units (OBUs), and the roadside units (RSUs), form a self-organized
Vehicular Ad Hoc Network (VANET)\cite{Bishop2000,Raya2007}. VANETs
inherently provide a way to collect traffic and road information
from vehicles, and to deliver road services including warnings and
traffic information to users in the vehicles.

Extensive research efforts have been made by both industry and
academia to investigate key issues in
VANETs\cite{Bishop2000,Mak2005,Xu2007}, with security and privacy
preservation as two primary
concerns\cite{Raya2007,Raya2005,Lin2007,Lin2008a,Lin2008b,Zhang2008a,Zhang2008b,Lu2008,Hubaux2004,Xi2007,Xi2008}.
Without security and privacy guarantees, attacks may jeopardize the
VANET's benefits: a malicious attack, such as a modification and
replay attack on the disseminated messages, could be fatal to some
users. Meanwhile, an attacker could trace the locations of the
vehicles and obtain their moving patterns if user-related private
information has not been protected. Hence, providing
privacy-preserving safety message\footnote{A safety message reports
on the state of the sender vehicle, e.g., its location, speed,
heading, etc.} authentication has become a fundamental design
requirement in securing VANETs.

The goals of privacy and liability are conflicting. On the one hand,
a well-behaved OBU is willing to offer as much local information as
possible to neighboring OBUs and RSUs to create a safer driving
environment on condition that its privacy has been well protected.
On the other hand, a malicious OBU may abuse the privacy protection
mechanism. This may particularly happen when a driver who is
involved in a dispute event of safety messages may attempt to avoid
legal responsibility. Therefore, the privacy-preserving message
authentication in VANETs should be conditional, such that a trusted
authority can disclose the real identity of targeted OBU in case of
a traffic event dispute, even though the OBU itself is not traceable
by the public.

The existing security and privacy solutions for VANETs can mainly be
categorized into three classes. The first one is based on a large
number of anonymous keys (denoted as LAB in the
following)\cite{Raya2005,Lin2008b}, the second one is based on a
pure group signature (denoted as GSB in the
following)\cite{Lin2007,Lin2008a}, while the last one employs the
RSU to assist vehicle in authenticating messages (denoted as RSUB in
the following)\cite{Zhang2008a,Zhang2008b,Lu2008}. Though all of
these solutions can meet the conditional privacy requirement, they
face obstacles in real deployments. First, the LAB scheme is not
efficient in terms of used storage and dispute solving. Second,
although the GSB scheme does not require each vehicle to store a
large number of anonymous keys, the time for message verification
grows linearly with the number of revoked vehicles. Worse, the
unrevoked vehicles have to update their private key and group public
key with the group manager when the number of revoked vehicles
surpasses some predefined threshold. This problem may be fatal for
VANET as they scale to cover all vehicles in a
country/continent.\footnote{At the moment, there are in the order of
some hundreds of millions of cars registered world wide}. Finally,
the RSUB protocol achieves much better efficiency than the previous
ones, however, the cost of deploying RSUs is high thus only some of
the roads can be covered especially at the initial deployment stage
of the VANET. Therefore, this solution may not be feasible in case
of the absence of the RSU.


To address these issues, this paper proposes an efficient and
spontaneous conditional privacy preservation protocol for
intervehicle communication based on revocable ring
signature\cite{Liu2007}. Compared to previous message-authentication
schemes\cite{Raya2007,Raya2005,Lin2007,Lin2008a,Lin2008b,Zhang2008a,Zhang2008b,Lu2008,Hubaux2004,Xi2007,Xi2008},
our scheme has the following unparalleled features: (1)
\textit{Conditional privacy}: Using the revocable ring signature to
secure the intervehicle communication, enables preserving privacy
regarding user identity and location of the vehicle, and the
identities of the target vehicles can be only revealed by the
trusted authority; (2) \textit{Efficiency}: The proposed protocol
can efficiently deal with a growing revocation list instead of
relying on a large storage space at each vehicle or updating the
group public key and private key at all unrevoked vehicles; (3)
\textit{Spontaneity}: The proposed protocol does not employ RSUs to
assist vehicles in authenticating messages while providing fast
message authentication and verification and an efficient conditional
privacy tracking mechanism. We believe this protocol is an excellent
candidate for the future VANETs.

The remainder of this paper is organized as follows. Section
\ref{sec2} surveys the related work. Section \ref{sec3} presents the
problem formulation, system architecture, and design objectives.
Section \ref{sec4} details the proposed security protocol, followed
by the security analysis and the performance analysis in Section
\ref{sec5} and Section \ref{sec6}, respectively. Section \ref{sec7}
concludes the paper.

\section{Background and Related Work}
\label{sec2}

\subsection{System Model}
\label{secIIA}

The considered system includes two types of entities: the
Transportation Regulation Center (TRC), and the moving vehicles
equipped with OBUs.

\begin{figure}[!t]
\centering
\includegraphics[scale=0.42]{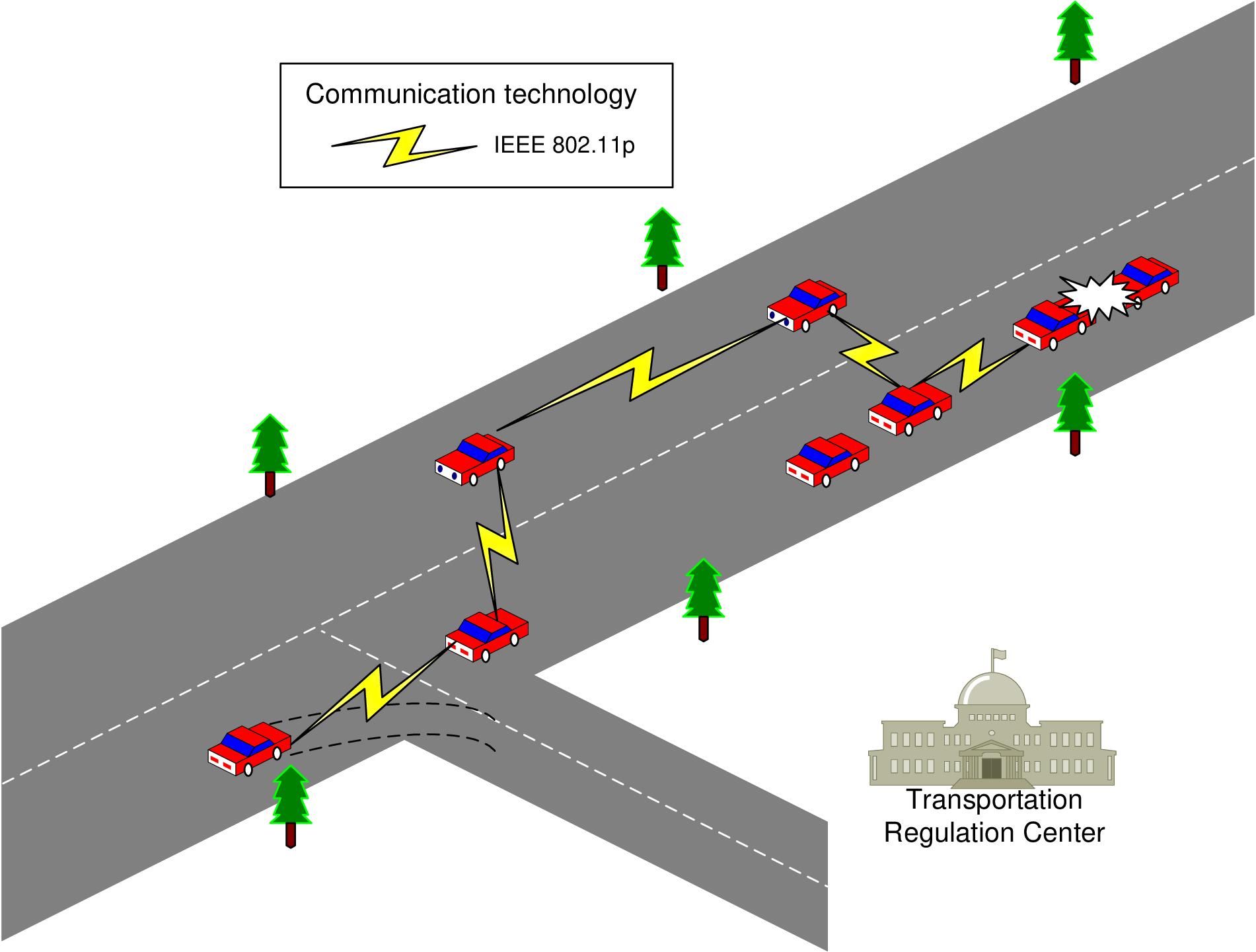}
\caption{System model: Road Emergency Operation under VANET}
\label{fig1}
\end{figure}

\begin{itemize*}
  \item OBU: All vehicles need to be registered with the TRC and
  preloaded with public system parameters and their own private key
  before the vehicle can join the VANETs.
  The use of secret information such as private keys generates the need for a
  tamper-proof device in each vehicle. The access to this device
  is restricted to authorized parties.
  OBUs are mobile and moving most of the time. When the OBUs are on
  the road, they regularly broadcast routine safety
  messages, such as position, current time, direction, speed,
  traffic conditions, traffic events, to help drivers get a
  better awareness of their environment
  and take early action to respond to an abnormal situation (Fig.
\ref{fig1}).
  Compared with the RSUs, the population of OBUs in the system could
  be up to millions, whereas the number of RSUs is at most tens of
  thousands based on the national infrastructure construction.
  \item TRC: TRC is in charge of the registration all OBUs each
vehicle is equipped with. The TRC can reveal the real identity of
  a safety message sender whenever there is a situation where the involved
  vehicles' IDs need to be revealed. The TRC has sufficient
  computation and storage capability, and is fully trusted by
  all parties in the system.
\end{itemize*}

Unlike other schemes, our solution does not employ RSUs. The network
dynamics are characterized by quasi-permanent mobility, high speeds,
and (in most cases) short connection times between neighbors. The
medium used for communications between neighboring OBUs is 5.9 GHz
Dedicated Short Range Communication (DSRC)\cite{DSRC} IEEE 802.11p.

\subsection{Related Work}

Many studies have been reported on the security and
privacy-preservation issues for
VANETs\cite{Raya2007,Raya2005,Lin2007,Lin2008a,Lin2008b,Zhang2008a,Zhang2008b,Lu2008,Hubaux2004,Xi2007,Xi2008}.
Xi \emph{et al.}\cite{Xi2007,Xi2008} introduced a random
key-set-based authentication protocol to preserve the vehicle's
privacy. However, they only provide the unconditional anonymity
without an effective and efficient tracking mechanism. To achieve
both message authentication and conditional anonymity, Raya \emph{et
al}.\cite{Raya2007,Raya2005} introduced a security protocol in
VANETs, namely LAB protocol, by requiring a large number of private
keys and corresponding anonymous certificates to be installed at
each vehicle. A vehicle randomly selects one of these anonymous
certificates and uses its corresponding private key to sign each
launched message. The other vehicles use the public key of the
sender enclosed in the anonymous certificate to authenticate the
source of the message. These anonymous certificates are generated by
employing the pseudo-identity of the vehicles, instead of taking any
real identity information of the drivers. Each certificate has a
short life time to meet the drivers'privacy requirement. Although
LAB protocol can effectively meet the conditional privacy
requirement, it is inefficient and may become a scalability
bottleneck. Because sufficient numbers of certificates must be
issued for each vehicle to maintain anonymity over a significant
period of time. As a result, the certificate database to be searched
by an authority in order to match a compromised certificate to its
owner's identity is huge.

Subsequently, Lin \emph{et al}.\cite{Lin2008b} developed a
time-efficient and secure vehicular communication scheme (TSVC)
based on the TESLA (Timed Efficient Stream Loss-tolerant
Authentication)\cite{Perrig2002}. With TSVC, a vehicle first
broadcasts a commitment of hash chain to its neighbors and then uses
the elements of the hash chain to generate a message authentication
code (MAC) with which other neighbors can authenticate this
vehicles' following messages. Because of the fast speed of MAC
verification, the communication and computation overhead of TSVC has
been reduced significantly. However, TSVC also requires a huge set
of anonymous public/private key pairs as well as their corresponding
public key certificates to be preloaded in each vehicle.
Furthermore, TSVC is not robust when the dynamics of traffic becomes
large since a vehicle should broadcast its key chain commitment much
more frequently.


Lin \emph{et al}.\cite{Lin2007,Lin2008a} proposed a security
protocol, i.e. GSB protocol, based on the group
signature\cite{Boneh2004}. With GSB, only a private key and the
group public key are stored in the vehicle, and the messages are
signed according to the group signature scheme without revealing any
identity information to the public. This assures that the trusted
authority is equipped with the capability of exposing the identity
of a sender. However, the time for safety message verification grows
linearly with the number of revoked vehicles in the revocation list.
Hence, each vehicle has to spend more time on safety message
verification. Furthermore, when the number of revoked vehicles in
the revocation list is larger than some threshold, the protocol
requires every remaining vehicle to calculate a new private key and
group public key based on the exhaustive list of revoked vehicles
whenever a vehicle is revoked. The means for system parameters to be
effectively updated to remaining vehicles, in a reliable and
scalable fashion, is not explored and represents an important
obstacle to the success of this scheme.

Recently, Zhang \emph{et al}.\cite{Zhang2008a,Zhang2008b} proposed a
novel RSU-aided message authentication scheme, that is RSUB, which
makes RSUs responsible for verifying the authenticity of messages
sent from vehicles and for notifying the results back to vehicles.
In this scheme, the vehicles have lower computation and
communication overhead than the previous reported schemes.
Independently, Lu \emph{et al}. \cite{Lu2008} introduced an
efficient conditional privacy preservation protocol in VANETs by the
generation of on-the-fly short-time anonymous keys between vehicles
and RSUs, which also can provide fast anonymous authentication and
privacy tracking. Both of these schemes explore an important feature
of VANETs by employing RSUs to assist vehicles in authenticating
messages. However, RSUs may not cover all the roads, especially in
the initial VANETs deployment stage, or due to the physical damage
of some RSUs, or simply for economic considerations.


\section{Preliminaries}
\label{sec3}

\subsection{Objectives}

To avoid reinventing the wheel, we refer the readers to other
works\cite{Lin2007,Raya2005} for a full discussion of the attacker
model. In the context of this work, we focus on the following
security objectives.
%

\begin{enumerate*}
  \item Efficient anonymous authentication of safety messages: The
  proposed scheme should provide an \textit{efficient} and
\textit{anonymous} message authentication mechanism. First, all
accepted messages should be
  delivered unaltered, and the origin of the messages should be authenticated to guard against impersonation attacks. Meanwhile,
  from the perspective of vehicle owners, it may not be acceptable to leak
  personal information, including identity and location, while
authenticating messages. Therefore, providing a
  secure yet anonymous message authentication is critical to the applicability of VANETs. Furthermore, the
  proposed scheme should be efficient in terms of fast verification on the safety messages and minimal anonymous
  keys storage at OBUs.
  \item Efficient tracking of the source of a disputed safety
  message: An important and challenging issue in these conditions is enabling TRC to retrieve a vehicle's real identity from its pseudo identity
  when a signature is in dispute or when the content of a message
  is bogus. Otherwise, anonymous authentication
  only can prevent an outside attack, but cannot deal with an inside
  one. That is to say, an insider can launch a bogus message
  spoofing attack or an impersonation attack successfully if the identity of the message sender can not be traced by the authorities.
  So it is necessary to provide traceability for the safety message to prevent the inside attack, otherwise concerns about the
  security may prevent vehicle owners from joining this system.
  \item Multilevel Anonymity\cite{Xi2007}: Privacy is a user-specific
  requirement and some users may be more serious about their privacy
  than others. Thus, it is noted that the proposed protocol should
  support multiple anonymity levels, and each vehicle should be allowed to choose
  its own anonymity level. The authentication protocol should
  provide a tradeoff between the anonymity level and resource
  utilization.
\end{enumerate*}

\subsection{Bilinear Maps}

Since bilinear maps\cite{Boneh2001} are the basis of our proposed
scheme, we briefly introduce them here.

Let $\mathbb{G}_{1}$ and $\mathbb{G}_{2}$ be two cyclic groups of
prime order $q$. Let $P$ be a generator of $\mathbb{G}_{1}$. Assume
that the discrete logarithm problem in both $\mathbb{G}_{1}$ and
$\mathbb{G}_{2}$ is hard. Suppose there exists a computable bilinear
map $\hat{e}$ such that $\hat{e}:\mathbb{G}_{1}\times
\mathbb{G}_{1}\rightarrow \mathbb{G}_{2}$ with the following
properties:

\begin{enumerate}
    \item Bilinearity: For all $P_{1},P_{2}\in \mathbb{G}_{1}$, and $a,b\in
\mathbb{Z}_{q}$,
$\hat{e}(aP_{1},bP_{2})=\hat{e}(P_{1},P_{2})^{ab}$.\item
Non-degeneracy: $\hat{e}(P,P)\neq 1_{\mathbb{G}_{2}}$.
\end{enumerate}

Such an admissible bilinear map $\hat{e}$ can be constructed by the
modified Weil or Tate pairing on elliptic curves. For example, the
Tate pairing on MNT curves\cite{Miyaji2001} gives the efficient
implementation, and the representations of $\mathbb{G}_{1}$ can be
expressed in $161$ bits when the order $q$ is a $160$-bit prime. By
this construction, the discrete logarithm problem in
$\mathbb{G}_{1}$ can reach $80$-bit security level.

\subsection{Ring Signature}

The ring signature scheme, introduced by Rivest, Shamir and
Tauman\cite{Rivest2001}, is characterized by two main properties:
anonymity and spontaneity. Anonymity in ring signature means
$1$-out-of-$n$ signer verifiability, which enables the signer to
keep anonymous in these ``rings" of diverse signers. Spontaneity is
a property which makes the distinction between ring signatures and
group signatures\cite{Chaum1991,Boneh2004}. Group signatures allow
the anonymity of a real signer in a group to be revoked by a trusted
party called group manager. It also gives the group manager the
absolute power of controlling the formation of the group. The ring
signature, on the other hand, does not allow anyone to revoke the
signer anonymity, while allowing the real signer to form a ring
arbitrarily without being controlled by any other party. Since
Rivest \textit{el al.}'s scheme, many ring signature schemes have
been
proposed\cite{Bresson2002,Abe2002,Wong2003,Boneh2003,Dodis2004}.

Recently, Liu et al.\cite{Liu2007} have introduced a new variant for
the ring signature, called revocable ring signature. This scheme
allows a real signer to form a ring arbitrarily while allowing a set
of authorities to revoke the anonymity of the real signer. In other
words, the real signer will be responsible for what has signed as
the anonymity is revocable by authorities while the real signer
still has the freedom on ring formation. We use this scheme as the
basis for our efficient and spontaneous conditional
privacy-preservation protocol.

\section{Efficient and Spontaneous Vehicular Communications Scheme}
\label{sec4}

This section describes in detail our efficient and spontaneous
privacy-preserving protocol for VANET. Each vehicle dynamically
collects the public keys of other vehicles it encounters during its
journey. Noted that this set of public keys keeps changing over
time. When the OBU wants to send a message, it uses these public
keys as its own group members to generate the revocable ring
signature. Furthermore, the identity of the sender can only be
recovered by the trusted authority.

The proposed scheme includes the following four phases: system
initialization, OBU safety message generation and sending, OBU
safety message verification, and OBU fast tracking algorithm. The
notation used throughout this paper is listed in Table \ref{tbl1}.

\begin{table}[h]\caption{Notations}
\label{tbl1}
\begin{tabular}{ll}\toprule
Notations  &  Descriptions \\
\hline
TRC:  &  \textbf{T}ransportation \textbf{R}egulation \textbf{C}enter \\
RL: & \textbf{R}evocation \textbf{L}ist\\
$V_{i}$:  &  The $i$th vehicle  \\
$\mathbb{G}_{1}$, $\mathbb{G}_{2}$: & two cyclic groups of
same order $q$\\
$P$: & The generator of $\mathbb{G}_{1}$\\
$RID_{i}:$ & The real identity of the vehicle $V_{i}$\\
$ID_{i}:$ & The pseudo-identity of the vehicle $V_{i}$\\
$M:$ & A message sent by the vehicle $V_{i}$\\
$x_{i}:$ & The private key of the vehicle $V_{i}$\\
$y_{i}=x_{i}P$: & The corresponding public key of the vehicle $V_{i}$\\
$x_{TRC}$: & The private key of the TRC\\
$y_{TRC}=x_{TRC}P$: & The corresponding public key of the TRC\\
$\mathcal{H}(\cdot):$ & A hash function such as $\mathcal{H}:\{0,1\}^{*}\rightarrow \mathbb{Z}_{q}$\\
$a\parallel b$ & String concatenation of $a$ and $b$\\
\bottomrule
\end{tabular}
\end{table}


\subsection{System Initialization}

Firstly, as described in section \ref{secIIA}, we assume each
vehicle is equipped with a tamper-proof device, which is secure
against any compromise attempt in any circumstance. With the
tamper-proof device on vehicles, an adversary cannot extract any
data stored in the device including key material, data, and codes
\cite{Raya2005,Hubaux2004}. We assume that there is a trusted
Transportation Regulation Center (TRC) which is in charge of
checking the vehicle's identity, and generating and pre-distributing
the private keys of the vehicles. Prior to the network deployment,
the TRC sets up the system parameters for each OBU as follows:

\begin{itemize*}
  \item Let $\mathbb{G}_{1}$, $\mathbb{G}_{2}$ be two cyclic groups of
same order $q$. Let $\hat{e}:\mathbb{G}_{1}\times
\mathbb{G}_{1}\rightarrow \mathbb{G}_{2}$ be a bilinear map.
  \item The TRC first randomly chooses $x_{TRC}\in_{R}\mathbb{Z}_{q}$ as its private key, and computes $y_{TRC}=x_{TRC}P$ as its public
  key. The TRC also chooses a secure cryptographic hash function
  $\mathcal{H}:\{0,1\}^{*}\rightarrow \mathbb{Z}_{q}$.
  \item The TRC generates a public and private key pair $(x_{i},y_{i})$ for each vehicle $V_{i}$ with real identity $RID_{i}$ as
  follows:
  By using $x_{TRC}$, the TRC first computes $x_{i}=\mathcal{H}(x_{TRC},RID_{i})\in\mathbb{Z}_{q}$, and then sets
  $y_{i}=x_{i}P\in\mathbb{G}_{1}$. In the end, the TRC stores the
  $(y_{i},RID_{i})$ in its records.
  \item Each vehicle is preloaded with the public parameters $\{\mathbb{G}_{1}, \mathbb{G}_{2}, q,
  y_{TRC}, \mathcal{H}\}$. In addition, the tamper-proof device of each vehicle is
  preloaded with its private/public key pairs $(x_{i},y_{i})$ and corresponding anonymous certificates
  (these certificates are generated by taking the vehicle's pseudo-identity $ID_{i}$). Finally, the vehicle will preload the revocation list (RL) from the
  TRC.
\end{itemize*}

\subsection{OBU Safety Message Generation}

%

Vehicle $V_{\pi}$ signs the message $M$ before sending it out.
Suppose $S=\{y_{1},\cdots,y_{n}\}$ is the set of public keys
collected by vehicle $V_{\pi}$ and it defines the ring of unrevoked
public keys. Note that the public key set $S$, collected and stored
temporarily by $V_{\pi}$, is dynamic. We assume that all public keys
$y_{i}$, $1\leq i \leq n$ and their corresponding private keys
$x_{i}$'s are generated by TRC, and $\pi$ ($1\leq \pi \leq n$) is
the index of the actual message sender. In other words, as $V_{\pi}$
travels through the road network, the set of public keys collected
by it keeps changing over time. Otherwise, a unique set of public
keys used by a vehicle may enable the adversary to infer its
traveling trajectory. The signature generation algorithm
$Sig(S,x_{\pi},y_{TRC},M)$ is carried out as follows.

\begin{enumerate*}
  \item Randomly select $r\in_{R}\mathbb{Z}_{q}$ and compute $R=rP$.
  \item For $y_{TRC}$, compute $E_{TRC}=\hat{e}(y_{\pi},y_{TRC})^{r}$.
  \item Generate a non-interactive proof $SPK(1)$ as follows: $SPK\{\alpha:\{E_{TRC}=\hat{e}(R,y_{TRC})^{\alpha}\}\bigwedge \{\bigvee\limits_{i\in [1,n]}y_{i}=\alpha
  P\}\}(M)$. The signature $\sigma$ of $M$ with respect to
  $S$ and $y_{TRC}$ is ($R,E_{TRC}$) and the transcript of $SPK(1)$.
\end{enumerate*}


For clear presentation, we divide $SPK(1)$ into two components:

\begin{subequations}
\begin{flalign}
\begin{split}
\indent SPK\{\alpha: E_{TRC}=\hat{e}(R,y_{TRC})^{\alpha}\}(M),
\end{split}&
\end{flalign}

\begin{flalign}
\begin{split}
\indent SPK\{\alpha: \bigvee\limits_{i\in [1,n]}y_{i}=\alpha
  P\}(M).
\end{split}&
\end{flalign}
\end{subequations}

To generate a transcript of $SPK(1a)$, given $E_{TRC},R,y_{TRC}$,
the actual message sender indexed by $\pi$ proves the knowledge of
$x_{\pi}$ such that $E_{TRC}=\hat{e}(R,y_{TRC})^{x_{\pi}}$ by
releasing $(s,c)$ as the transcript such that
$$c=\mathcal{H}(y_{TRC}\parallel R\parallel E_{TRC}\parallel \hat{e}(R,y_{TRC})^{s}E_{TRC}^{c}\parallel M)$$

This can be done by randomly picking $l\in_{R}\mathbb{Z}_{q}$ and
computing $$c=\mathcal{H}(y_{TRC}\parallel R\parallel
E_{TRC}\parallel \hat{e}(R,y_{TRC})^{l}\parallel M)$$ and then
setting $s=l-cx_{\pi}\bmod q$.

To generate the transcript of $SPK(1b)$, given $S$, the actual
message sender indexed by $\pi$, for some $1\leq \pi \leq n$, proves
the knowledge of $x_{\pi}$ out of $n$ discrete logarithms $x_{i}$,
where $y_{i}=x_{i}P$, for $1\leq i \leq n$, without revealing the
value of $\pi$. This can be done by releasing
$(s_{1},\cdots,s_{n},c_{1},\cdots,c_{n})$ as the transcript such
that $c_{0}=\sum_{i=1}^{n}c_{i}\bmod q$ and
$$c_{0}=\mathcal{H}(S\parallel s_{1}P+c_{1}y_{1}\parallel \cdots \parallel s_{n}P+c_{n}y_{n}\parallel
M).$$

To generate this transcript, the actual message sender first picks
randomly $l\in_{R}\mathbb{Z}_{q}$ and
$s_{i},c_{i}\in_{R}\mathbb{Z}_{q}$ for $1\leq i \leq n$, $i\neq
\pi$, then computes
\begin{eqnarray*}
c_{0} &=& \mathcal{H}(S\parallel s_{1}P+c_{1}y_{1}\parallel \cdots\parallel s_{\pi-1}P+c_{\pi-1}y_{\pi-1}\parallel lP\parallel \\
      && s_{\pi+1}P+c_{\pi+1}y_{\pi+1}\parallel \cdots\parallel
      s_{n}P+c_{n}y_{n}\parallel M)
\end{eqnarray*}

and finds $c_{\pi}$ such that $c_{0}=c_{1}+\cdots+c_{n}\bmod q$.
Finally the actual message sender sets
$s_{\pi}=l-c_{\pi}x_{\pi}\bmod q$.

Now we combine the constructions of $SPK(1a)$ and $SPK(1b)$
together. First, the actual message sender randomly picks
$l_{1},l_{2}\in_{R}\mathbb{Z}_{q}$ and
$s_{i},c_{i}\in_{R}\mathbb{Z}_{q}$ for $1\leq i \leq n$, $i\neq
\pi$, then computes
\begin{eqnarray*}
c &=& \mathcal{H}(S\parallel y_{TRC}\parallel R\parallel
E_{TRC}\parallel \hat{e}(R,y_{TRC})^{l_{1}}\parallel\\
      && s_{1}P+c_{1}y_{1}\parallel \cdots\parallel s_{\pi-1}P+c_{\pi-1}y_{\pi-1}\parallel l_{2}P\parallel \\
      && s_{\pi+1}P+c_{\pi+1}y_{\pi+1}\parallel \cdots\parallel
      s_{n}P+c_{n}y_{n}\parallel M).
\end{eqnarray*}

After that, the actual message sender sets $s=l_{1}-cx_{\pi}\bmod
q$, finds $c_{\pi}$ such that $c=c_{1}+\cdots+c_{n}\bmod q$, and
sets $s_{\pi}=l_{2}-c_{\pi}x_{\pi}\bmod q$. The transcript of
$SPK(1)$ is therefore $(s,s_{1},\cdots,s_{n},c_{1},\cdots,c_{n})$.



\subsection{Message Verification}
\label{4-3}

Once a message is received, the receiving vehicle first checks if
the $RL\bigcap S\buildrel?\over=\emptyset$. If so, the receiver
performs signature verification by verifying of $SPK(1)$ as follows:

\begin{eqnarray*}
\sum_{i=1}^{n}c_{i} &\buildrel?\over=& \mathcal{H}(S\parallel
y_{TRC}\parallel R\parallel
E_{TRC}\parallel \\
      && \hat{e}(R,y_{TRC})^{s}E_{TRC}^{\sum_{i=1}^{n}c_{i}}\parallel s_{1}P+c_{1}y_{1}\parallel
      \\
      && \cdots\parallel s_{n}P+c_{n}y_{n}\parallel
      M).
\end{eqnarray*}

After that, the receiving vehicle updates its own public key set by
randomly choosing public keys from $S$.

\subsection{OBU fast tracing}
\label{secIVD}

A membership tracing operation is performed when solving a dispute,
where the real ID of the signature generator is desired. The TRC
first checks the validity of the signature and then uses its private
key $x_{TRC}$ and determines if
$$E_{TRC}\buildrel?\over=\hat{e}(y_{i},R)^{x_{TRC}}$$ for some $i$, $1\leq i\leq n$.

If the equation holds at, say when $i=\pi$, then the TRC looks up
the record $(y_{\pi},RID_{\pi})$ to find the corresponding identity
$RID_{\pi}$ meaning that vehicle with identity $RID_{\pi}$ is the
actual message generator. The TRC then broadcasts the
$(y_{\pi},RID_{\pi})$ to all OBUs and each OBU adds the $y_{\pi}$
into his local revocation list (RL).

\section{Security Analysis}
\label{sec5}

We analyze the security of the proposed scheme in terms of the
following four aspects: message authentication, user identity
privacy preservation, traceability by the TRC, and spontaneity of
the signature generator.

\begin{itemize*}
  \item \emph{Message authentication}. Message authentication is
  the basic security requirement in vehicular
  communications. In the proposed scheme, the ring signature can
  only be generated by the valid ring members. Without knowing any
  of the discrete logarithms $x_{i}$ of the public keys $y_{i}$ in
  the ring
  $S$, it is infeasible to forge a valid ring signature.
  \item \emph{Identity privacy preservation}. Given a valid ring
  signature $\sigma$ of some message, it is computationally
  difficult to identify the actual signer by any participant in the system except the TRC.
  If there exists an algorithm which breaks the signer anonymity of
  the construction in Section \ref{sec4}, then the
  Indistinguishability Based Bilinear Decisional Diffie-Hellman
  assumption would be contradicted\cite{Liu2007}.
  \item \emph{Traceability}. Given the signature, only the TRC who knows $x_{TRC}$,
  can trace the real identity of a message sender using the OBU tracking procedure described in section \ref{secIVD}. Besides, the
  tracing process carried by the TRC does not require any
  interaction with the message generator. Instead, the revocable ring signature itself provides the authorship information to TRC. Therefore, once a signature is in
  dispute, the TRC has the ability to trace the disputed message, in
  which the traceability can be well satisfied.
  \item \emph{Spontaneity}. Note that the actual message generator
  can specify the ring (a set of vehicles) required to generate the ring signature arbitrarily
  based on the public keys of vehicles it encountered in the past without any new interaction with
  any other vehicles or RSUs in the system. Compared with the schemes
  \cite{Zhang2008a,Zhang2008b,Lu2008}, our scheme does not use the RSUs to
  assist vehicles in authenticating messages.
  \item \emph{Multilevel privacy.} Each vehicle can
  select the degree of privacy that fits its own requirements by
  choosing the number of public keys used in the message
  generation phase. This way, each vehicle can achieve a proper balance between privacy
  protection and resource usage. The multilevel privacy of our
  scheme gives users flexibility in defining their
  privacy requirements.
\end{itemize*}



\section{Performance Evaluation}
\label{sec6}

This section evaluates the performance of the proposed scheme in
terms of storage requirements and computational overheads.

\subsection{Storage Overheads}

This subsection compares the OBU storage overhead of our protocol
with three previously proposed protocols: LAB\cite{Raya2005},
RSUB\cite{Lu2008} and GSB\cite{Lin2007}. In the LAB protocol, each
OBU stores not only its own $N_{okey}$ anonymous key pairs, but also
all the anonymous public keys and their certificates in the
revocation list (the notations adopted in the description are listed
in Table \ref{tbl3}). Let each key (with its certificate) occupy one
storage unit. If there are $m$ OBUs revoked, then the scale of
revoked anonymous public keys is $m\cdot N_{okey}$. Thus, the total
storage overhead in LAB protocol (denoted as $S_{LAB}$) is
$S_{LAB}=(m+1)N_{okey}$. Assuming that $N_{okey}=10^{4}$, we have
$S_{LAB}=(m+1)10^{4}$. Both in our protocol and GSB protocol, each
OBU stores one private key issued by the trusted party, and $m$
revoked public keys in the revocation list. Let $S_{GSB}$ and
$S_{RRSB}$ denote the total storage unit of GSB protocol and our
protocol (Revocable Ring Signature Based protocol) respectively.
Thus, $S_{GSB}=S_{RRSB}=m+1$. In the RSUB protocol\cite{Lu2008},
each OBU stores one private key issued by the trusted party, and a
short-time key pair together with its anonymous certificate issued
by the RSU. Since the OBU does not need to store the revocation
list, the storage overhead in RSUB protocol is only two units,
denoted as $S_{RSUB}=2$.



\begin{figure}[!t]
\centering
\includegraphics[width=3.6in]{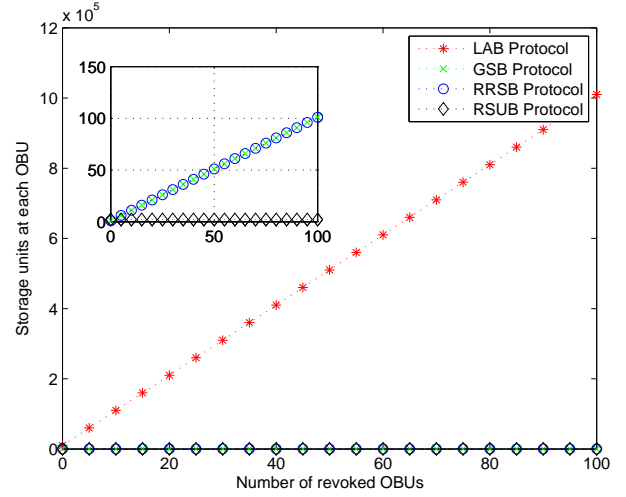}
\caption{Each OBU storage overhead of Raya's, Lin's, Lu's and our
protocol in different $m$ revoked OBUs, $m$ varying from 1 to 100}
\label{fig2}
\end{figure}

Fig.\ref{fig2} shows the storage units of LAB protocol, GSB
protocol, RSUB protocol and our protocol as $m$ increases. Observe
that the OBU storage overhead in LAB protocol linearly increases
with $m$, and is much larger than that in the other three protocols.
The storage overhead of GSB protocol and our protocol is still small
in spite of its linear increase with $m$. Though the storage
overhead in RSUB protocol is the most efficient, this scheme
requires the RSUs, instead of OBUs, to store the anonymous key
pairs, which, nonetheless, is not the case in the other schemes.

\begin{table}[h]\caption{Notations and Rough Scale}
\label{tbl3}
\begin{tabular}{lll}\toprule
&  Descriptions & Scale\\
\hline
$N_{obu}:$  &  The number of OBUs in the system & $10^{7}$ \\
$N_{okey}:$  &  The number of anonymous keys owned by one OBU & $10^{4}$ \\
$N_{rsu}:$  &  The number of RSUs in the system & $10^{4}$ \\
$N_{rkey}:$  &  The number of anonymous keys processed by one RSU & $10^{4}$ \\
\bottomrule
\end{tabular}
\end{table}

\begin{table}[h]\caption{Cryptography Operation's Execution Time}
\label{tbl4}
\begin{tabular}{lll}\toprule
&  Descriptions & Execution Time\\
\hline
$T_{pmul}$  &  The time for one point multiplication in $\mathbb{G}$ & 0.6 ms \\
$T_{pair}$  &  The time for one pairing operation & 4.5 ms \\
\bottomrule
\end{tabular}
\end{table}

\subsection{Message Verification Overhead}

This subsection compares the OBU computation overhead for the
proposed, RSUB and GSB protocols. Since the point multiplication in
$\mathbb{G}$ and pairing computations dominates each party's
computation overhead, we consider only these operations in the
following estimation. Table \ref{tbl4} gives the measured processing
time (in milliseconds) for an MNT curve\cite{Miyaji2001} of
embedding degree $k=6$ and 160-bit $q$. The implementation was
executed on an Intel pentium IV 3.0 GHz machine\cite{Scott2007}.

In our proposed protocol, verifying a message, requires
$T_{pair}+(2n+1)T_{pmul}$, where $n$ is the cardinality of the ring,
as shown in section \ref{4-3}. Let $T_{RRSB}$ be the required time
cost in our protocol, then we have:
$$T_{RRSB}=T_{pair}+(2n+1)T_{pmul}=4.5+(2n+1)\times0.6 (ms)$$

In the GSB protocol, the time cost to verify a message is related to
the number of revoked OBUs in the revocation list. Thus the required
time is:
$$T_{GSB}=6T_{pmul}+(4+m)T_{pair}=6\times0.6+(4+m)\times4.5 (ms)$$

In the RSUB protocol, to verify a message, it requires
$3T_{pair}+11T_{pmul}$. Let $T_{RSUB}$ be the required time cost in
RSUB's protocol, then we have:
$$T_{RSUB}=3T_{pair}+11T_{pmul}=3\times 4.5+11\times0.6=20.1 (ms)$$

\begin{figure}[!t]
\centering
\includegraphics[width=3.6in]{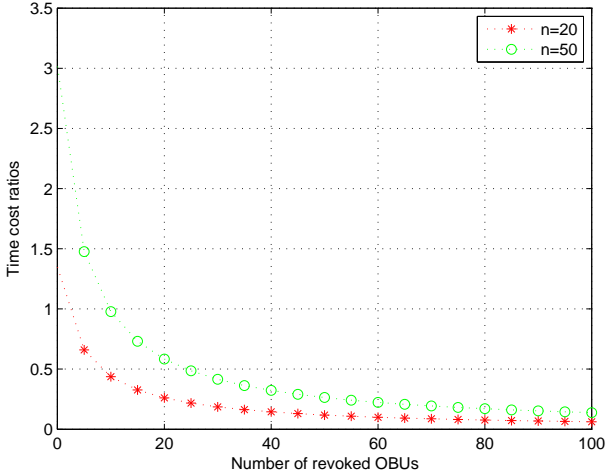}
\caption{Time efficiency ratio $T_{RG}=T_{RRSB}/T_{GSB}$ with a
number of $m$ revoked OBUs, $m$ varying from 1 to 100.} \label{fig3}
\end{figure}

Let
$$T_{RG}=\frac{T_{RRSB}}{T_{GSB}}$$
be the cost ratio between our proposed protocol and the GSB
protocol. Fig.\ref{fig3} plots the time cost ratio $T_{RG}$ when $m$
OBUs are revoked, as $m$ ranges from 1 to 100. We observe that the
time cost ratio $T_{RG}$ decreases as $m$ increases, which
demonstrates the much better efficiency of our proposed protocol
than the GSB protocol especially when the revocation list is large.
Note that $n$ can be determined by the user according to its own
computation capacity and privacy requirements.

\begin{figure}[!t]
\centering
\includegraphics[width=3.6in]{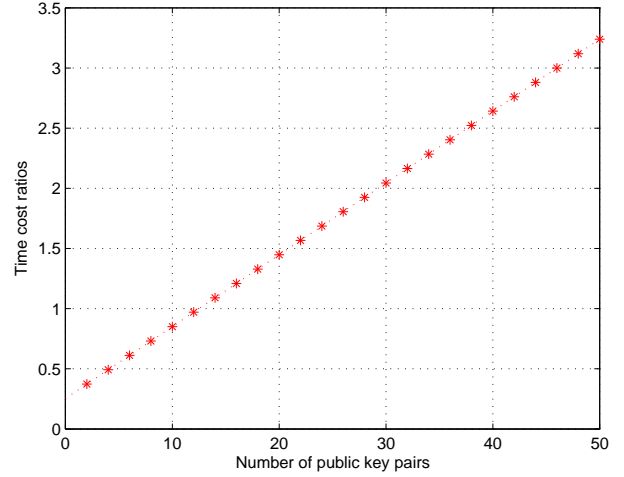}
\caption{Time efficiency ratio $T_{RR}=T_{RRSB}/T_{RSUB}$ with a
number of $n$ public key pairs, $n$ varying from 1 to 50.}
\label{fig4}
\end{figure}

Let
$$T_{RR}=\frac{T_{RRSB}}{T_{RSUB}}$$
be the cost ratio between our proposed protocol and RSUB protocol.
Fig.\ref{fig4} plots the time cost ratio $T_{RR}$ when $n$ public
key pairs are employed, where the number of $n$ ranges from 1 to 50.
We observe that the time cost ratio $T_{RR}$ increases as $n$
increases, which demonstrates our protocol is slightly more
expensive than RSUB protocol. However, our protocol does not employ
the roadside infrastructures to communicate with the OBU as in RSUB
protocol, which will cause additional communication overhead.

\subsection{Trusted Authority Computation Complexity on OBU Tracing}

In this subsection, we evaluate the trusted authority computation
complexity on OBU tracing algorithm. For fair comparison, we use the
same linear and binary search algorithms in all of these protocols.
We use the same notations as in the previous sections. Table
\ref{tbl5} presents the computation complexity for the four
protocols. The trusted authority tracking algorithm in our proposed
protocol and GSB protocol has the better efficiency than the other
two protocols.

\begin{table}[h]\caption{Comparison of Computation Complexity}
\label{tbl5}
\begin{tabular}{lll}\toprule
Protocol &  Linear search & Binary search\\
\hline
LAB:  &  $O(N_{obu}\cdot N_{okey})$ & $O(log(N_{obu}\cdot N_{okey}))$ \\
GSB:  &  $O(N_{obu})$ & $O(log(N_{obu}))$ \\
RSUB: & $O(N_{rsu}+N_{rkey})$ & $O(log(N_{rsu}\cdot N_{rkey}))$ \\
RRSB:  &  $O(N_{obu})$ & $O(log(N_{obu}))$ \\
\bottomrule
\end{tabular}
\end{table}

\section{Summary}
\label{sec7}

We have presented an efficient, spontaneous, conditional privacy
preserving protocol based on the revocable ring signature and aimed
for secure vehicular communications. We demonstrate that proposed
protocol is not only provides conditional privacy, a critical
requirement in VANETs, but also able to improve efficiency in terms
of the number of keys stored at each vehicle, identity tracking in
case of a dispute, and, most importantly message authentication and
verification. Meanwhile, our proposed solution can operate
independently: does not require support from the roadside
infrastructure which, at least in the initial deployment stages, may
not cover all road segments.



\bibliographystyle{ieeetr}

\begin{thebibliography}{1}


\bibitem{IVI} Saving Lives Through Advanced Vehicle Safety Technology: Intelligent Vehicle Initiative
Final Report. [Online]. Available:
http://www.itsdocs.fhwa.dot.gov/JPODOCS/REPTS\rule[-1pt]{0.12cm}{0.2pt}PR/14153\rule[-1pt]{0.12cm}{0.2pt}files/ivi.pdf

\bibitem{VII} Vehicle infrastructure integration. U.S. Department of
Transportation, [Online]. Available:
http://www.its.dot.gov/index.htm

\bibitem{DoT2006} U.S. Department of Transportation, National
Highway Traffic Safety Administration, \emph{Vehicle Safety
Communications Project}, Final Report. Appendix H: WAVE/DSRC
Security, April 2006.

\bibitem{Misener2005} J.~A.~Misener, ``Vehicle-infrastructure integration (VII) and safety",
\emph{Intellimotion}, Vol. 11, No. 2, pp. 1-3, 2005.

\bibitem{Bishop2000} R.~Bishop, ``A survey of intelligent vehicle applications
worldwide", in \emph{Proceedings of the IEEE Intelligent Vehicles
Symposium 2000}, Dearborn, MI, USA, Oct. pp. 25-30, 2000.

\bibitem{Mak2005} T. K.~Mak, K.~P.~Laberteaux and R.~Sengupta, ``A Multi-Channel
VANET Providing Concurrent Safety and Commercial Services," in
\emph{Proceedings of 2nd ACM International Workshop on Vehicular Ad
Hoc Networks}, Cologne, Germany, Sep. pp. 1-9, 2005.

\bibitem{Xu2007} Q.~Xu, T.~Mak, J.~Ko and R.~Sengupta, ``Medium Access Control
Protocol Design for Vehicle-Vehicle Safety Messages," \emph{IEEE
Transactions on Vehicular Technology}, March 2007, Vol. 56, No. 2,
pp. 499-518.

\bibitem{Xi2007} Y.~Xi, K.~Sha, W.~Shi, L.~Scnwiebert, and T.~Zhang, ``Enforcing
Privacy Using Symmetric Random Key-Set in Vehicular Networks",
\emph{Eighth International Symposium on Autonomous Decentralized
Systems (ISADS'07)}, pp. 344-351, 2007.

\bibitem{Xi2008} Y.~Xi, W.~Shi, L.~Schwiebert. ``Mobile anonymity of dynamic
groups in vehicular networks", \emph{Security and Communication
Networks}, Vol. 1, No.3, pp. 219-231, 2008.

\bibitem{Raya2007} M.~Raya and J.~P.~Hubaux, ``Securing Vehicular Ad Hoc Networks",
\emph{Journal of Computer Security}, Special Issue on Security of Ad
Hoc and Sensor Networks, Vol. 15, Nr. 1, pp. 39 - 68, 2007.

\bibitem{Raya2005} M.~Raya, J.~P.~Hubaux, ``The security of vehicular ad hoc networks",
\emph{3rd ACM workshop on Security of ad hoc and sensor networks},
2005, pp. 11-21.

\bibitem{Lin2007} X.~Lin, X.~Sun, P.-H.~Ho and X.~Shen. ``GSIS: A Secure
and Privacy-Preserving Protocol for Vehicular Communications",
\emph{IEEE Transactions on Vehicular Technology}, vol. 56(6), pp.
3442-3456, 2007.

\bibitem{Lin2008a} X.~Lin, R.~Lu, C.~Zhang, H.~Zhu, P.-H.~Ho and X.~Shen. ``Security in
Vehicular Ad Hoc Networks", \emph{IEEE Communications Magazine},
vol. 46, no. 4, pp. 88-95, 2008.

\bibitem{Lin2008b} X.~Lin, X.~Sun, X.~Wang, C.~Zhang, P.-H.~Ho and X.~Shen, ``TSVC:
Timed Efficient and Secure Vehicular Communications with Privacy
Preserving", \emph{IEEE Transactions on Wireless Communications},
vol. 7, no. 12, pp. 4987-4998, 2008.

\bibitem{Zhang2008a} C.~Zhang, X.~Lin, R.~Lu and P.-H.~Ho. RAISE: An Efficient
RSU-aided Message Authentication Scheme in Vehicular Communication
Networks. IEEE International Conference on Communications (ICC'08),
Beijing, China, May 19-23, 2008.

\bibitem{Zhang2008b} C.~Zhang, X.~Lin, R.~Lu, P.-H.~Ho and X.~Shen. ``An Efficient Message
Authentication Scheme for Vehicular Communications", \emph{IEEE
Transactions on Vehicular Technology}, vol. 57, no. 6, pp.
3357-3368, 2008.

\bibitem{Lu2008} R.~Lu, X.~Lin, H.~Zhu, P.-H.~Ho and X.~Shen. ``ECPP: Efficient
Conditional Privacy Preservation Protocol for Secure Vehicular
Communications", \emph{The 27th IEEE International Conference on
Computer Communications (INFOCOM 2008)}, Phoenix, Arizona, USA,
April 15-17, 2008.

\bibitem{Hubaux2004} J.P.~Hubaux, S.~Capkun, L.~Jun, ``The Security and Privacy of Smart
Vehicles", \emph{IEEE Security \& Privacy Magazine}, Vol. 2, No. 3,
pp. 49-55, 2004.

\bibitem{Calandriello2007} G.~Calandriello, P.~Papadimitratos, J.-P.~Hubaux, A.~Lioy,
``Efficient and robust pseudonymous authentication in VANET",
\emph{Vehicular Ad Hoc Networks} pp. 19-28, 2007.

\bibitem{Dötzer2005} F.~Dötzer, ``Privacy Issues in Vehicular Ad Hoc Networks",
\emph{Privacy Enhancing Technologies}, pp. 197-209, 2005.

\bibitem{DSRC} \emph{Dedicated Short Range Communications} (5.9 GHz DSRC), Available: http://www.leearmstrong.com/DSRC/DSRCHomeset.htm.

\bibitem{Perrig2002} A.~Perrig, R.~Canetti, J.~D.~Tygar, D.~Song, ``The TESLA Broadcast Authentication
Protocol", RSA CryptoBytes, vol. 5, no. 2, pp. 2-13, 2002.

\bibitem{Boneh2001} D.~Boneh and M.~K.~Franklin, ``Identity-based encryption
from the Weil pairing," In \emph{J. Kilian, editor, CRYPTO 2001},
volume 2139 of LNCS, pages 213-229. Springer, 2001.

\bibitem{Miyaji2001} A.~Miyaji, M.~Nakabayashi, and S.~Takano, ``New explicit
conditions of elliptic curve traces for FR-reduction," \emph{IEICE
Transactions on Fundamentals}, Vol.E84-A, No.5, pp.1234-1243, 2001.

\bibitem{Scott2007} M.~Scott, ``Efficient Implementation of Cryptographic pairings", [on-line]. Availabe: http://ecrypt-ss07.rhul.ac.uk/Slides/Thursday/mscott-samos07.pdf

\bibitem{Rivest2001} R.~L.~Rivest, A.~Shamir, Y.~Tauman, ``How to Leak a Secret", In
\emph{AsiaCrypt 2001}, volume 2248 of LNCS, pp. 552-565.

\bibitem{Chaum1991} D.~Chaum, E.~van Hevst, ``Group Signature", In \emph{EUROCRYPT
1991},volume 547 of LNCS, pp. 257-265.

\bibitem{Boneh2004} D.~Boneh, X.~Boyen, H.~Shacham, ``Short group signatures", In:
Franklin, M.K. (ed.) CRYPTO 2004. vol 3152 of LNCS, pp. 41–55,
Springer, Heidelberg (2004).

\bibitem{Bresson2002} E.~Bresson, J.~Stern, M.~Szydlo, ``Threshold ring signatures and applications to ad-hoc groups,"
In \emph{Proc. CRYPTO 2002}, USA, Lecture Notes in Computer Science,
2442, Springer-Verlag, 2002, pp.465~480.


\bibitem{Abe2002} M.~Abe, M.~Ohkubo, K.~Suzuki, ``1-out-of-n signatures from a variety
of keys," In \emph{Proc. ASIACRYPT 2002}, New Zealand, Lecture Notes
in Computer Science, 2501, Springer-Verlag, 2002, pp.415~432.


\bibitem{Wong2003} D.~S.~Wong, K.~Fung, J.~Liu, V.~Wei, ``On the RS-code construction of
ring signature schemes and a threshold setting of RST," In
\emph{Proc. 5th Int. Conference on Infoation and Communication
Security (ICICS 2003)}, China, Lecture Notes in Computer Science,
2836, Springer-Verlag, 2003, pp.34~46.


\bibitem{Boneh2003} D.~Boneh, C.~Gentry, B.~Lynn, H.~Shacham, ``Aggregate and verifiably
encrypted signatures from bilinear maps," In \emph{Proc. EUROCRYPT
2003}, Poland, Lecture Notes in Computer Science, 2656,
Springer-Verlag, 2003, pp.416~432.


\bibitem{Dodis2004} Y.~Dodis, A.~Kiayias, A.~Nicolosi, V.~Shoup, ``Anonymous identification
in ad doc groups," In \emph{Proc. EUROCRYPT 2004}, Switzerland, LNCS
3027, Springer-Verlag, 2004, pp.609~626, Full version:
http://www.cs.nyu.edu/~nico-lo-si/pa-pers/

\bibitem{Liu2007} D.~Y.~W.~Liu, J.~K.~Liu, Y.~Mu, W.~Susilo, D.S.~Wong: ``Revocable
Ring Signature", \emph{J. Comput. Sci. Technol.} 22(6): pp. 785-794.
2007.

\end{thebibliography}
\setlength{\baselineskip}{13pt}
\end{document}